# Superconducting single photon detectors integrated with diamond nanophotonic circuits

Patrik Rath[1], Oliver Kahl[1], Simone Ferrari[1], Fabian Sproll[2], Georgia Lewes-Malandrakis[3], Dietmar Brink[3], Konstantin Ilin[2], Michael Siegel[2], Christoph Nebel[3], and Wolfram Pernice[1,*]

[1] *Institute of Nanotechnology, Karlsruhe Institute of Technology, 76021 Karlsruhe, Germany*

[2] *Institute of Micro- and Nanoelectronic Systems, Karlsruhe Institute of Technology, 76187 Karlsruhe, Germany*

[3] *Fraunhofer Institute for Applied Solid State Physics, Tullastr. 72, 79108 Freiburg, Germany*

* Corresponding author electronic mail: wolfram.pernice@kit.edu

## Abstract

Photonic quantum technologies promise to repeat the success of integrated nanophotonic circuits in non-classical applications. Using linear optical elements, quantum optical computations can be performed with integrated optical circuits and thus allow for overcoming existing limitations in terms of scalability. Besides passive optical devices for realizing photonic quantum gates, active elements such as single photon sources and single photon detectors are essential ingredients for future optical quantum circuits. Material systems which allow for the monolithic integration of all components are particularly attractive, including III-V semiconductors, silicon and also diamond. Here we demonstrate nanophotonic integrated circuits made from high quality polycrystalline diamond thin films in combination with on-chip single photon detectors. Using superconducting nanowires coupled evanescently to travelling waves we achieve high detection efficiencies up to 66 % combined with low dark count rates and timing resolution of 190 ps. Our devices are fully scalable and hold promise for functional diamond photonic quantum devices.

# Introduction

The advent of integrated optical circuits has led to a surge of applications in telecommunications, optical signal processing and on-chip sensing[1–4]. Combining individually optimized photonic components into complex systems provides a flexible framework for the realization of compact and powerful devices. While such circuits are predominantly investigated in classical optics, nanophotonic devices hold further promise for emerging applications in quantum optics and photonic quantum technologies[5,6]. Ever since the realization that optical quantum computing is possible using linear optical elements, single photon sources and detectors, the realization of these three essential elements in a single circuit has been a driving goal[7–9]. A particular requirement in this respect is the scalability of all core elements. While scalability can be readily achieved for linear optical elements through nanofabrication and miniaturization, scalable implementation of the active parts of the circuits is less straightforward. Here we demonstrate the scalable realization of single photon detectors embedded in diamond nanophotonic circuits as a first step towards fully integrated diamond quantum systems.

Single photon detectors which are fast and provide high detection efficiency combined with good timing accuracy are highly sought after for applications in quantum optics, imaging, as well as metrology and sensing. Depending on the wavelength regime of interest, different technologies can be explored such as silicon avalanche photodiodes (APD) for visible wavelengths[10], photomultiplier tubes or InGaAs based APDs for the telecommunication range[11]. In the telecoms range APDs provide non-perfect performance because of relatively high dark count rates and the need for gated-mode operation. Furthermore, the integration with nanophotonic circuitry is non-trivial. In recent years superconducting nanowire single photon detectors (SNSPDs)[12–14] have been shown to be promising alternatives, especially when integrated directly onto waveguides and into photonic circuits[15–17]. In order to take advantage of the immense optical bandwidth of SNSPDs[18] which allows one to detect single photons in the mid infrared[19], at the important near infrared wavelengths[20] (telecom C-band), as well as in the visible spectrum[21], a substrate material for waveguide integrated SNSPDs is needed which is transparent throughout this broad wavelength range. Furthermore, the substrate needs to be available as a high quality thin film in order to support guided modes in waveguides.

Diamond with its wide electronic bandgap of 5.47 eV is emerging as a promising material for integrated optics because of its broadband transparency spectrum from 225 nm into the far infrared. Additionally diamond provides a relatively high refractive index of 2.4 which allows for tight confinement of light into subwavelength waveguides. By making use of chemical vapor deposition diamond thin films on a wafer-scale have become available as a convenient template for nanophotonic circuit fabrication[22,23]. Besides applications in sensing and optomechanics [24–29], diamond is also of broad interest for quantum optics[30–32]. While waveguide integrated SNSPDs have been shown on wide-bandgap silicon nitride[16,33], realizing integrated single photon detectors on diamond nanophotonic circuitry is of special importance, as diamond hosts color centers suitable for the realization of efficient single-photon sources, such as the nitrogen vacancy center[34–37] and the silicon vacancy center[31,38–40].

Here we present integrated single photon detectors fabricated atop high quality polycrystalline diamond nanophotonic waveguides. Using a travelling wave layout we are able to achieve single photon detection on-chip with high efficiency. Our detectors have a minimal footprint allowing us to fabricate hundreds of detector circuits on a single chip. We show that the travelling wave detector can be readily integrated into diamond nanophotonic circuits in a scalable fashion, using established lithography and dry etching procedures. Our results provide a promising route towards all-diamond single photon circuits for on-chip quantum optics.

## Materials and Methods

**Detector design and device fabrication**

In order to combine diamond nanophotonic circuits with superconducting single photon detectors, we fabricate niobium nitride (NbN) nanowires directly on top of diamond waveguides, as shown in the schematic illustration in Fig. 1a. The NbN nanowires are connected to larger metal contact pads, as shown in the scanning electron microscopy (SEM) image in Fig. 1b, which allow us to electrically connect the nanowires to a bias current source and appropriate readout electronics. This way, single photons which are absorbed in the NbN nanowire can be efficiently detected. Due to the small footprint of the devices, hundreds of photonic circuits, each equipped with their own single photon detector, can be fabricated in one fabrication run on a 15x15 mm$^2$ wafer die, as shown on the microscope image Fig.1c.

Opposed to earlier implementations of SNSPDs, where photons impinge onto the NbN nanowires from the top[12], here a so called "travelling wave" geometry[15] is implemented, meaning that photons propagate along a waveguide and travel parallel along the NbN nanowire. In this design the propagating photons are coupled evanescently to the superconducting nanowire and are absorbed in the optical near field. Therefore the absorption length can be arbitrarily increased by extending the wire length, in contrast to the classical SNSPD geometry where photons impinge onto the detector under normal incidence from the optical far field[12]. In the latter case the absorption length is limited to the thickness of the NbN layer which is only a few nanometers thick. Travelling wave SNSPDs, on the other hand, feature absorption lengths of tens of micrometers, leading to absorption efficiencies which approach 100 %[15], thus overcoming a crucial limitation of traditional SNSPDs. Our waveguide detectors are embedded in nanophotonic circuits as shown in Fig.1c. The circuits make use of focusing grating couplers[41,42] for launching light into the on-chip waveguides and extracting transmitted light at a second coupling port. The incoming light is split at a 50:50 beam splitter for calibration of the light intensity inside the waveguide. Light propagating along a second path is guided to the detector which is attached to the waveguide leading upwards, away from the grating couplers.

In order to realize diamond-based photonic circuitry, diamond thin films surrounded by cladding materials with lower refractive index are required. For this purpose a 1 µm thick polycrystalline diamond film is deposited by plasma enhanced chemical vapor deposition (PECVD)[43,44] onto an oxidized silicon wafer with 2 µm of $SiO_2$ on top of a silicon carrier wafer, resulting in a wafer-scale diamond-on-insulator (DOI) template[22,28]. Subsequently the diamond layer is polished to a thickness of 600 nm by chemo mechanical polishing with a soft cloth[45–47] to reduce the root mean square surface roughness below 3 nm[46]. Finally, a NbN thin film is sputter deposited onto the diamond-on-insulator (DOI) wafer using DC reactive magnetron sputtering in an argon and nitrogen gas mixture. The resulting NbN layer has a thickness of 4.6 nm and a critical temperature of 6.5 K, well above the base temperature of 1.8K of the cryostat used in this work. While the diamond thin films used in this work are polycrystalline the fabrication and design routines are general. Therefore transferring our approach to single crystal diamond-on-insulator substrates is possible. Recent progress in realizing single crystalline DOI templates via transfer techniques provide promising steps in this direction[48–50]. For compatibility with our waferscale fabrication approach, however, here we restrict our work to high quality microcrystalline diamond thin films.

The photonic circuits and the waveguide integrated SNSPDs are fabricated using three steps of electron beam lithography (EBL) with a JEOL 5500 system at 50 kV acceleration voltage and subsequent dry etching steps. During the first EBL step electrodes and alignment markers are written into PMMA positive tone resist. Then 5 nm of chromium, 150 nm of gold and finally 10 nm of chromium are deposited via electron beam physical vapor deposition (PVD) under ultra-high vacuum conditions. A subsequent lift-off step in acetone finalizes the metal structures. A thin glass layer of 5 nm is afterwards sputtered onto the NbN film for resist adhesion promotion. Negative tone resist HSQ 6% is spin coated with a thickness of 120 nm and structured via EBL to define the nanowires and connection pads to the gold electrodes. This allows us to write thin wires of widths below 100 nm, while providing a robust etch mask during dry etching using argon plasma to remove the glass adhesion layer and $CF_4$ chemistry to fully etch the NbN layer. Fig.1d shows an atomic force microscopy (AFM) image of such a patterned nanowire, allowing us to quantify the nanowire dimensions, as well as remaining surface roughness of the DOI wafer.

Next, another glass layer of 5 nm is sputtered onto the now exposed diamond layer for adhesion promotion during the final EBL step. The photonic structures are defined using 480 nm thick HSQ 15% and transferred into the diamond thin film by reactive ion etching in argon and oxygen plasma[47]. Etching away 300 nm of the initial diamond layer results in half etched rib waveguides as illustrated in Fig.1a. In the resulting overall structure the NbN wire is protected from the environment by a HSQ layer which remains on top of the SNSPD and the waveguide.

**Devices for the measurement of the absorption efficiency**

Besides dedicated single photon detector circuits, the full chip also contains additional photonic circuits to characterize the absorption properties of the NbN nanowires. The absorption efficiency of light propagating inside the waveguide increases with increasing nanowire width because of a larger waveguide coverage with superconducting material. For broadband detection of single photons, however, narrow nanowires are desirable because the smallest photon energy for which photons can be detected decreases with increasing nanowire width. Therefore the geometry of the SNSPDs has to be optimized for a given wavelength and waveguide mode. In particular, for detecting single photons in the telecommunication range the nanowire width should be less than 150 nm. To increase the absorption efficiency with narrow nanowires it is therefore preferred to use a meander layout instead of wider nanowires. Because of the large number of photonic circuits on each chip we are able to investigate multiple detector device geometries. In particular, two general designs are fabricated and studied in this work, as depicted in Fig.2a: first a single meander geometry, consisting of two parallel straight sections and one connecting bend, and second a double meander geometry, consisting of four parallel straight sections connected in series.

We simulate the guided modes for a 1 μm wide half etched waveguide for 1550 nm input wavelength with finite element methods using COMSOL Multiphysics. Fig. 2b shows the mode profiles and corresponding effective refractive indices for a waveguide with a single meander (I and II), with a double meander nanowire (III) and without NbN nanowire (IV). The waveguide dimensions were intentionally chosen such that the waveguide supports only one transverse electric (TE)-like mode. Due to the large complex refractive index of NbN the electric field at the nanowire is greatly enhanced, as can be seen in the close-up image (II), which leads to the desired enhanced absorption of photons in the NbN nanowire.

Back reflection of photons at the interface between a bare waveguide and a waveguide with a NbN wire on top is a possible limiting factor for the detection efficiency of a detector. We estimate the upper bound of this reflection by assuming a sudden change from the effective refractive index of the guided mode between the bare waveguide ($n_{eff}$ = 2.128) and the waveguide with a double meander of 100nm nanowire width ($n_{eff}$ = 2.130 - 0.0086i), where the imaginary part corresponds to an absorption of -0.302 dB/µm. Using Fresnel's equations we find a reflection coefficient $R = \frac{I_r}{I_i} = 4.5 * 10^{-6}$, where $I_r$ and $I_i$ are the reflected and the incident intensity respectively. This very low reflection value can be safely neglected in the following analysis.

In order to experimentally determine the absorption efficiency of the NbN nanowires on chip we fabricate dedicated photonic circuits which allow for balanced optical detection[15]. These devices are fabricated on the same chip and in the same fashion as the detector devices presented in Fig. 1. The balanced detection design, as illustrated in Fig. 2a, consists of the following components: a focusing grating coupler, as shown in the inset of Fig. 2a, designed for efficient coupling of light from a fiber to the TE-like waveguide mode at 1550 nm, is used to launch light into the input waveguide. A Y-Splitter is used as a 50:50 beam splitter: half of the light is guided to the right side and coupled out of the chip via a second grating coupler, which acts as reference port. The other half is guided to the left side, where the propagating mode is attenuated by absorption in the NbN nanowire and afterwards coupled out at the "transmission port" grating coupler. Each nanowire is parameterized by a certain wire width and wire length which is varied from device to device across the chip. This balanced detection design hence allows us to determine the attenuation due to absorption by the NbN by dividing the transmitted power $P_{trans}$ by the power at the reference port $P_{ref}$. By using the reference port with identical optical grating couplers and waveguides, propagation loss and coupling loss induced into the circuit do not contribute to the measurement of the absorption coefficient.

## Results and Discussion

**Measurement of the detector absorption efficiency**

Fabricated samples as described in the previous section are placed on a fiber-optic measurement setup and aligned against an optical fiber array. Light from a tunable infrared (IR) laser source (New Focus TLB6600) is coupled into the central waveguide while the transmitted signals are recorded at both output ports simultaneously. The attenuation due to the nanowire can thus be extracted by calculating the ratio of the measured transmitted intensities at the transmission and the reference ports. We measure the attenuation for nanowires of three different widths (75 nm, 100 nm, 125 nm), as shown in Fig. 2c for the double meander geometry. The measurements are initially performed at room temperature. The measured attenuation shows the expected exponential decay with increasing wire length. We extract the attenuation in dB/μm from linear fits to the obtained data for double meanders (Fig. 2c) and single meanders (not shown). The fit results, depicted in Fig. 2d, show the expected attenuation increase with wire width and with the number of parallel NbN nanowires per waveguide. The largest measured attenuation, for 125 nm wide double meanders amounts to -0.278 dB/μm. This is slightly smaller than predicted by the numerical simulations, which we attribute to the fabricated NbN wire being slightly narrower than the designed width.

**Cryogenic measurement of the detector speed**

For detector characterization at cryogenic temperatures, the detector chip is placed into a liquid helium flow cryostat, as illustrated in Fig. 3 and cooled down to a base temperature of 1.8K. The chip is mounted on a stack of nano-positioners with closed-loop feedback (Attocube Systems) for alignment against the fiber array. The piezo stage allows for controlling the position of the chip in three dimensions and further enables in-plane rotation. At room temperature the chip is prealigned against the fiber array and the positions of all detector devices are recorded. After cool down, the photonic circuits are first aligned with respect to the fiber array for light insertion and collection, by moving the nano-positioners in the x/y-plane (see coordinate system in Fig. 3). Then the RF probe is brought into contact with the gold contact pads of the devices for electrical connection of the detectors, by moving the nano-positioners in the z-direction. The superconducting nanowires are current biased using a low-noise current source and a bias-T acting as low pass filter. They are furthermore connected to the readout electronics through a second GHz bias-T in order to separate the high-frequency SNSPD signal from the DC bias.

The optical setup consists either of a tunable IR continuous wave laser or a pulsed IR laser combined with two variable optical attenuators and a polarization controller. The on-chip reference port signal is monitored with a lightwave multimeter to enable precise control of the photon flux reaching the SNSPDs with the adjustable optical attenuators which provide up to 60 dB attenuation each. The SNSPD are both biased and read-out via the connected RF probe. The collected signal is then electrically amplified by 85 dB using low-noise RF amplifiers and eventually recorded on a single photon counting system or a fast oscilloscope. High signal-to-noise ratio of the photon detection event is achieved by choosing low noise electrical amplifiers with appropriate bandwidth, gain and noise-figure.

Fig. 4a shows the average of 500 pulses from a SNSPD measured with a high speed oscilloscope (blue). The decay time of the pulse is extracted from an exponential fit (red). For a 35 µm long double meander detector the fit yields a decay time of 5.1 ns, corresponding to a potential maximum detector count rate of about 200 MHz, which compares favorably to state-of-the-art photomultiplier tubes and avalanche photo-diodes.[13]

**Determination of the on-chip detection efficiency**

We then measure the on-chip detection efficiency (OCDE) of our detectors by comparing the photon flux arriving at the detector with the detector count rate in dependence of bias current. The dark count rate is determined at the same certain bias current but with the laser turned off. Fig. 4b illustrates the detection efficiency measurement: each photonic circuit consists of an input grating coupler, which launches light from the fiber coupled laser into the diamond waveguide. After a subsequent 50:50 Y-splitter, half of the light is routed to the SNSPD, while the other half of the light is routed to an output grating coupler, which acts as a reference port. Taking into account the grating coupler, the 50:50 splitter and the propagation loss, the rate of photons arriving at the detector is determined. We then employ two optical attenuators to attenuate the light, such that the photon flux arriving at the detector is around $50000\,{}^{1}/_{s}$, which is far below the maximum detector count rate.

The OCDE is calculated from the measured count rate CR, corrected for the dark count rate DCR, and the photon flux $\Phi$ arriving at the detector: $OCDE = \frac{CR-DCR}{\Phi}$. The photon flux is calculated using the on-chip reference port. For this, we first measure the transmitted power at the reference port $P_{out}$ for a known laser power $P_{in}$. The input and output coupling efficiency are considered to be the same. We calculate the coupling efficiency C under the current measurement condition as $C = \sqrt{\frac{P_{out}}{P_{in} \times S \times WGL_1}}$, where S = 0.5 is the splitting ratio of the Y-splitter and WGL₁ is the propagation loss along the waveguide connecting the input and output coupler.

To assess the propagation loss for half etched diamond rib waveguides of 1 μm width a chip with half etched ring resonators is fabricated following the same fabrication steps as for the detector chip. For weakly coupled ring resonators of 1 μm width the average quality factor accounts to Q = 9650, which corresponds to a propagation loss of $\alpha = 4.79\ dB/mm$. The photon flux Φ arriving at the SNSPD is calculated as $\Phi = P_{in,att} \times C \times S \times WGL_2$, with the attenuated laser power $P_{in,att}$, coupling efficiency C and the propagation loss $WGL_2$ along the waveguide connecting the Y-splitter and the SNSPD. The on-chip detection efficiency (OCDE) hence follows as $OCDE = \frac{CR-DCR}{P_{in,att} \times (C \times S) \times WGL_2}$.

The count rate measurements are carried out with an integration time of 2 s. The uncertainty in the count rates is estimated from the variance of ten consecutive measurements and (except for very low count rates) found to be about 1 %. To account for statistical variations in both grating coupler efficiency C and splitter ratio S we evaluate reference port transmissions $T = C^2 \times S$ of the balanced splitter devices shown in Fig. 2a. The transmission of 54 devices of identical geometry is measured to be $T = 3.44 \times 10^{-3} \pm 3.78 \times 10^{-4}$, which is equivalent to a relative variation in transmission of $\frac{\Delta T}{T} = \frac{\Delta(C^2 \times S \times WGL_1)}{C^2 \times S \times WGL_1} = 10.98\%$. For the OCDE calculation one coupler and one splitter contribute to the photon flux. An upper bound for the uncertainty in $C \times S$ is hence given by $\frac{\Delta(C \times S)}{C \times S} \leq \frac{\Delta(C^2 \times S \times WGL_1)}{C^2 \times S \times WGL_1} = 10.98\%$. Eventually, the uncertainty of the propagation loss is calculated from the variation in quality factors extracted from the resonances of a ring resonator. For the studied ring resonators we consistently observe a variation in Q factor of 5%. Error propagation leads to a relative uncertainty in propagation loss of $\frac{\Delta WGL_2}{WGL_2}$ =1.57%.

As expected, when increasing the bias current $I_{bias}$ the on-chip detection efficiency increases as shown in Fig. 4c. The bias current is given relative to the critical current of a nanowire. For a 100 nm wide SNSPD the measured critical current at 1.9K is 3.4 µA, corresponding to a critical current density of $j_C = 0.74 \frac{MA}{cm^2}$. Several detectors with identical wire width, but increasing wire length from 35 µm to 80 µm are compared in this graph. For detectors with longer wires the on-chip detection efficiency improves due to the increase in absorption efficiency.[51] In the case of an 80 µm long single meander detector the measured on-chip detection efficiency is 40% when biased at 96% of the critical current. As expected from the absorption measurements, detectors with the same device length but different meander conformation show different efficiencies (Fig. 4d). Hence the double meanders have a higher efficiency than the single meanders. For a 65 µm long double meander detector the measured OCDE accounts to 66%. This means that for 100 photons arriving at the detector on average 66 detection events are recorded, showing that at the applied operation conditions the energy of one infrared photon is sufficient to break the superconductivity and lead to a measureable output signal. Variations in measured OCDE between devices of the same geometry are attributed to the residual surface roughness of the diamond layer, which can lead to constrictions in the wire, limiting the maximum possible bias current and hence the detector efficiency.[52] For the efficiency measurements the photon flux arriving at the detector is far below the saturation count rate of the SNSPDs. The detectors are operated at very low light intensities, such that the absorption of two photons in the same place and at the same time is very improbable. We furthermore confirm that our SNSPDs operate as single photon detectors by confirming that the count rate scales linearly with the attenuated laser power, when biased close to the critical current.[53,54] For this type of detector this unambiguously demonstrates the single-photon detection capability.[12]

**Determination of dark count rates and timing accuracy**

In addition to the OCDE, we analyze the dark count rate (DCR) for every device at the same conditions which were used during the efficiency measurements. For this purpose the laser is turned off, but the fiber array stays mounted on top of the grating couplers. Fig.5a shows the dark counts measured for the 65 µm long double meander with 66% OCDE (see Fig. 4d). When increasing the bias current the dark count rate rises exponentially[55] to a maximum value of 1.8 kHz, when biased at 96% of the critical current. When the fiber array is mounted on top of the grating couplers, the dark count rate is typically limited by stray light, which is coupled into the cryostat via the optical fibers[16]. We calculate the on-chip noise equivalent power (NEP$_{OC}$) which is defined as $NEP_{OC} = h\nu \times \frac{\sqrt{2DCR}}{OCDE}$, where $h\nu$ is the photon energy. The lowest NEP for this device occurs when operating the detector at 81% of its critical current. The detector then shows on-chip detection efficiency of 38%, while the dark count rate is below 3 Hz. The minimum noise equivalent power therefore amounts to $NEP_{OC} = 7.9 \times 10^{-19} \frac{W}{\sqrt{Hz}}$.

We then analyze the timing accuracy of the SNSPDs by estimating their jitter contribution using a fast oscilloscope in histogram mode. A picosecond pulsed fiber laser at 1550 nm (Pritel) is used with a repetition rate of 40 MHz, providing ~1 ps pulses with a timing jitter below 1 ps. The laser light passes through a 50:50 splitter and the resulting equal portions are routed to a fast low noise photo-receiver (1GHz New Focus 1611) and to the SNSPDs in the cryostat. The SNSPDs are operated close to their critical current and the detection events are recorded with a fast 6 GHz digital oscilloscope (Agilent 54855A). The signal of the photo-receiver acts as the trigger signal. The jitter of the oscilloscope and the 1 GHz photodetector have been measured to be less than 1 picosecond, respectively, well below the measured detection jitter. Therefore the bandwidth limit and the total timing jitter are mainly determined by the low noise amplifiers and the SNSPD.

During the jitter measurements the SNSPD signal is triggered at half the voltage pulse amplitude where the maximum slope is reached. The SNSPD is operated close to its critical current and the measurement is performed using different sets of amplifiers as shown in Fig. 5b. Gaussian fits of the data reveal a FWHM jitter value of 186 ps. This value is strongly influenced by the electrical instrumentation as shown in Fig.5b and can be interpreted as an upper limit of the detector's true timing jitter.

## Conclusions

Using the approach outlined above we successfully show the realization of two key elements for on-chip quantum optics in diamond: efficient on-chip single photon detectors combined with integrated photonic circuits. Our detectors are fast, with a decay time of 5.1 ns, enabling a maximum detector count rate up to 200 MHz. The devices also provide on-chip detection efficiencies as high as 66% and noise equivalent powers as low as $7.9 * 10^{-19} \frac{W}{\sqrt{Hz}}$. Because the on-chip detection efficiency improves for shorter wavelengths[19,21,33], we expect that the detection efficiency for photons at visible wavelengths will be even higher and could be further increased with improved polishing procedures. Furthermore, the fabrication and design approach is general and can be directly transferred to single crystalline diamond-on-insulator substrates.

The implementation of travelling wave SNSPDs on diamond is a promising step towards a quantum-optics-on-a-chip platform which relies on monolithically joining single photon sources, single photon routing and processing devices, as well as single photon detectors. We envision that this could be achieved by combining optical cavities of high quality factor [36,40,56–58] with color centers as single photon sources and long waveguides for photon routing[22,59]. Furthermore, by utilizing mechanical degrees of freedom available in free-standing diamond waveguides, tunable elements and phase shifters[28,29,60] can be realized. By detecting these photons with fast and efficient single photon detectors on the same chip, a full suite of functional elements is thus available for diamond quantum photonics.

## Acknowledgements

W.H.P. Pernice acknowledges support by the DFG grants PE 1832/1-1 & PE 1832/1-2 and the Helmholtz society through grant HIRG-0005. P. Rath acknowledges financial support by the Deutsche Telekom Stiftung. The PhD education of P. Rath and O. Kahl is embedded in the Karlsruhe School of Optics & Photonics (KSOP). We also acknowledge support by the Deutsche Forschungsgemeinschaft (DFG) and the State of Baden-Württemberg through the DFG-Center for Functional Nanostructures (CFN) within subproject A6.4. We thank S. Kühn and S. Diewald for the help with device fabrication and F. Pyatkov for help with initial resistivity measurements. The authors furthermore thank F. Wu, M. Blaicher and M. Stegmaier for helpful discussions concerning measurement setup and manuscript.

# References:

1. Soref, R. The Past, Present, and Future of Silicon Photonics. *IEEE Journal of Selected Topics in Quantum Electronics* **12,** 1678–1687 (2006).

2. Koos, C. *et al.* All-optical high-speed signal processing with silicon–organic hybrid slot waveguides. *Nature Photonics* **3,** 216–219 (2009).

3. Robinson, J. T., Chen, L. & Lipson, M. On-chip gas detection in silicon optical microcavities. *Optics Express* **16,** 4296 (2008).

4. Vollmer, F. & Arnold, S. Whispering-gallery-mode biosensing: label-free detection down to single molecules. *Nature methods* **5,** 591–6 (2008).

5. O'Brien, J. L., Furusawa, A. & Vučković, J. Photonic quantum technologies. *Nature Photonics* **3,** 687–695 (2009).

6. Duan, L.-M. & Kimble, H. Scalable Photonic Quantum Computation through Cavity-Assisted Interactions. *Physical Review Letters* **92,** 127902 (2004).

7. Knill, E., Laflamme, R. & Milburn, G. J. A scheme for efficient quantum computation with linear optics. *Nature* **409,** 46–52 (2001).

8. Kok, P., Nemoto, K., Ralph, T. C., Dowling, J. P. & Milburn, G. J. Linear optical quantum computing with photonic qubits. *Reviews of Modern Physics* **79,** 135–174 (2007).

9. O'Brien, J. L. Optical quantum computing. *Science* **318,** 1567–70 (2007).

10. Dautet, H. *et al.* Photon counting techniques with silicon avalanche photodiodes. *Applied optics* **32,** 3894–900 (1993).

11. Itzler, M. A. *et al.* Advances in InGaAsP-based avalanche diode single photon detectors. *Journal of Modern Optics* **58,** 174–200 (2011).

12. Gol'tsman, G. N. *et al.* Picosecond superconducting single-photon optical detector. *Applied Physics Letters* **79,** 705 (2001).

13. Hadfield, R. H. Single-photon detectors for optical quantum information applications. *Nature Photonics* **3,** 696–705 (2009).

14. Natarajan, C. M., Tanner, M. G. & Hadfield, R. H. Superconducting nanowire single-photon detectors: physics and applications. *Superconductor Science and Technology* **25,** 063001 (2012).

15. Pernice, W. H. P. *et al.* High-speed and high-efficiency travelling wave single-photon detectors embedded in nanophotonic circuits. *Nature communications* **3,** 1325 (2012).

16. Schuck, C., Pernice, W. H. P. & Tang, H. X. Waveguide integrated low noise NbTiN nanowire single-photon detectors with milli-Hz dark count rate. *Scientific reports* **3,** 1893 (2013).

17. Schuck, C., Pernice, W. H. P., Ma, X. & Tang, H. X. Optical time domain reflectometry with low noise waveguide-coupled superconducting nanowire single-photon detectors. *Applied Physics Letters* **102,** 191104 (2013).

18. Henrich, D. *Influence of Material and Geometry on the Performance of Superconducting Nanowire Single-Photon Detectors*. (KIT Scientific Publishing, 2013).

19. Marsili, F. *et al.* Efficient single photon detection from 500 nm to 5 μm wavelength. *Nano letters* **12,** 4799–804 (2012).

20. Marsili, F. *et al.* Detecting single infrared photons with 93% system efficiency. *Nature Photonics* **7,** 210–214 (2013).

21. Verevkin, A. *et al.* Detection efficiency of large-active-area NbN single-photon superconducting detectors in the ultraviolet to near-infrared range. *Applied Physics Letters* **80,** 4687 (2002).

22. Rath, P. *et al.* Waferscale nanophotonic circuits made from diamond-on-insulator substrates. *Optics Express* **21,** 11031 (2013).

23. Checoury, X. *et al.* Nanocrystalline diamond photonics platform with high quality factor photonic crystal cavities. *Applied Physics Letters* **101,** 171115 (2012).

24. Kolkowitz, S. *et al.* Coherent sensing of a mechanical resonator with a single-spin qubit. *Science* **335,** 1603–6 (2012).

25. Neumann, P. *et al.* High precision nano scale temperature sensing using single defects in diamond. *Nano letters* (2013). doi:10.1021/nl401216y

26. Cai, J., Jelezko, F. & Plenio, M. B. Hybrid sensors based on colour centres in diamond and piezoactive layers. *Nature communications* **5,** 4065 (2014).

27. Teissier, J., Barfuss, A., Appel, P., Neu, E. & Maletinsky, P. Strain Coupling of a Nitrogen-Vacancy Center Spin to a Diamond Mechanical Oscillator. *Physical Review Letters* **113,** 020503 (2014).

28. Rath, P., Khasminskaya, S., Nebel, C., Wild, C. & Pernice, W. H. P. Diamond-integrated optomechanical circuits. *Nature Communications* **4,** 1690 (2013).

29. Rath, P. *et al.* Diamond electro-optomechanical resonators integrated in nanophotonic circuits. *Applied Physics Letters* **105,** 251102 (2014).

30. Dolde, F. *et al.* Room-temperature entanglement between single defect spins in diamond. *Nature Physics* **9,** 139–143 (2013).

31. Rogers, L. J. *et al.* Multiple intrinsically identical single-photon emitters in the solid state. *Nature communications* **5,** 4739 (2014).

32. Aharonovich, I. & Neu, E. Diamond Nanophotonics. *Advanced Optical Materials* **2,** 911–928 (2014).

33. Schuck, C., Pernice, W. H. P. & Tang, H. X. NbTiN superconducting nanowire detectors for visible and telecom wavelengths single photon counting on Si3N4 photonic circuits. *Applied Physics Letters* **102,** 051101 (2013).

34. Englund, D. *et al.* Deterministic coupling of a single nitrogen vacancy center to a photonic crystal cavity. *Nano letters* **10,** 3922–6 (2010).

35. Faraon, A., Santori, C., Huang, Z., Acosta, V. & Beausoleil, R. Coupling of Nitrogen-Vacancy Centers to Photonic Crystal Cavities in Monocrystalline Diamond. *Physical Review Letters* **109,** 033604 (2012).

36. Hausmann, B. J. M. *et al.* Coupling of NV Centers to Photonic Crystal Nanobeams in Diamond. *Nano letters* **13,** 5791–6 (2013).

37. Doherty, M. W. *et al.* The nitrogen-vacancy colour centre in diamond. *Physics Reports* **528,** 1–45 (2013).

38. Lee, J. C., Aharonovich, I., Magyar, A. P., Rol, F. & Hu, E. L. Coupling of silicon-vacancy centers to a single crystal diamond cavity. *Optics express* **20,** 8891–7 (2012).

39. Neu, E. *et al.* Low-temperature investigations of single silicon vacancy colour centres in diamond. *New Journal of Physics* **15,** 043005 (2013).

40. Riedrich-Möller, J., Arend, C. & Pauly, C. Deterministic coupling of a single silicon-vacancy color center to a photonic crystal cavity in diamond. *Nano letters* **14,** 5281–5287 (2014).

41. Taillaert, D. *et al.* An out-of-plane grating coupler for efficient butt-coupling between compact planar waveguides and single-mode fibers. *IEEE Journal of Quantum Electronics* **38,** 949–955 (2002).

42. Rath, P., Khasminskaya, S., Nebel, C., Wild, C. & Pernice, W. H. P. Grating-assisted coupling to nanophotonic circuits in microcrystalline diamond thin films. *Beilstein Journal of Nanotechnology* **4,** 300–305 (2013).

43. Fuener, M., Wild, C. & Koidl, P. Novel microwave plasma reactor for diamond synthesis. *Applied Physics Letters* **72,** 1149 (1998).

44. Fuener, M., Koidl, P. & Wild, C. Simulation and development of optimized microwave plasma reactors for diamond deposition. *Surface and Coatings Technology* **116-119,** 853–862 (1999).

45. Thomas, E. L. H., Nelson, G. W., Mandal, S., Foord, J. S. & Williams, O. A. Chemical mechanical polishing of thin film diamond. *Carbon* **68,** 473–479 (2014).

46. Ummethala, S. *et al.* High-Q optomechanical circuits made from polished nanocrystalline diamond thin films. *Diamond and Related Materials* **44,** 49–53 (2014).

47. Rath, P. *et al.* Diamond Nanophotonic Circuits Functionalized by Dip-pen Nanolithography. *Advanced Optical Materials* (2014). doi:10.1002/adom.201400434

48. Ovartchaiyapong, P., Pascal, L. M. A., Myers, B. A., Lauria, P. & Bleszynski Jayich, A. C. High quality factor single-crystal diamond mechanical resonators. *Applied Physics Letters* **101,** 163505 (2012).

49. Tao, Y. & Degen, C. Facile fabrication of single-crystal-diamond nanostructures with ultrahigh aspect ratio. *Advanced materials* **25,** 3962–7 (2013).

50. Lee, J. C., Magyar, A. P., Bracher, D. O., Aharonovich, I. & Hu, E. L. Fabrication of thin diamond membranes for photonic applications. *Diamond and Related Materials* **33,** 45–48 (2013).

51. Kovalyuk, V. *et al.* Absorption engineering of NbN nanowires deposited on silicon nitride nanophotonic circuits. *Optics express* **21,** 22683–92 (2013).

52. Kerman, A. J. *et al.* Constriction-limited detection efficiency of superconducting nanowire single-photon detectors. *Applied Physics Letters* **90,** 101110 (2007).

53. Elezov, M. S. *et al.* Investigating the detection regimes of a superconducting single-photon detector. *Journal of Optical Technology* **80,** 435 (2013).

54. Ferrari, S. *et al.* Waveguide-integrated single- and multi-photon detection at telecom wavelengths using superconducting nanowires. *Applied Physics Letters* **106,** 151101 (2015).

55. Kitaygorsky, J. *et al.* Origin of Dark Counts in Nanostructured NbN Single-Photon Detectors. *IEEE Transactions on Appiled Superconductivity* **15,** 545–548 (2005).

56. Bayn, I. *et al.* Fabrication of triangular nanobeam waveguide networks in bulk diamond using single-crystal silicon hard masks. *Applied Physics Letters* **105,** 211101 (2014).

57. Hausmann, B. J. M. *et al.* Integrated High Quality Factor Optical Resonators in Diamond. *Nano letters* **13,** 1898–1902 (2013).

58. Lee, J. C. *et al.* Deterministic coupling of delta-doped nitrogen vacancy centers to a nanobeam photonic crystal cavity. *Applied Physics Letters* **105,** 261101 (2014).

59. Hiscocks, M. P. *et al.* Diamond waveguides fabricated by reactive ion etching. *Optics express* **16,** 19512–9 (2008).

60. Sohn, Y.-I., Burek, M. J. & Lončar, M. Dynamic Actuation of Single-Crystal Diamond Nanobeams. *arXiv preprint:1408.5822* (2014). at <http://arxiv.org/abs/1408.5822>

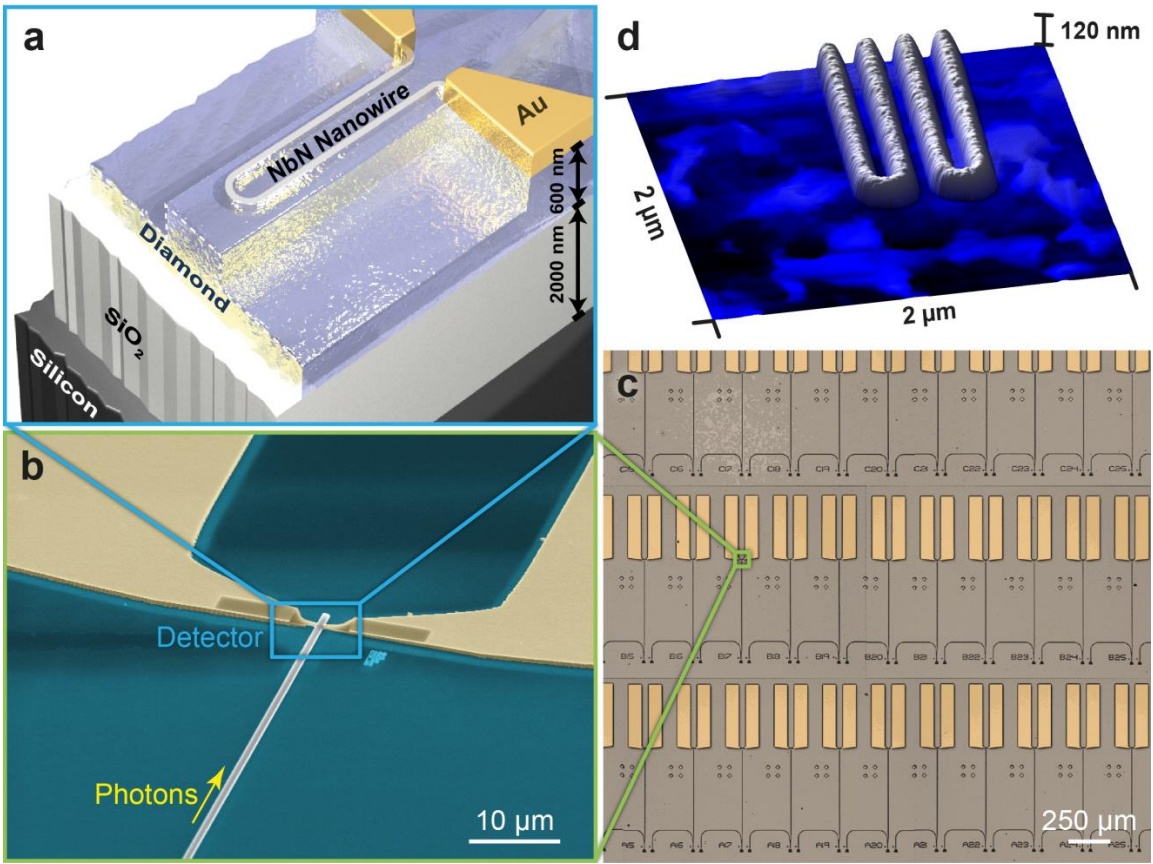

**Figure 1. Diamond nanophotonic circuits containing superconducting nanowire detectors.**
a) Schematic of a NbN superconducting nanowire detector on top of a diamond rib waveguide. Gold contact pads enable electrical contact with the detector. b) False color SEM image of one waveguide integrated detector connected to the gold contact pads. c) Optical microscope image showing a selection of the 192 detectors which were fabricated per sample. d) Atomic force microscopy scan over a double meander nanowire after the first lithography step.

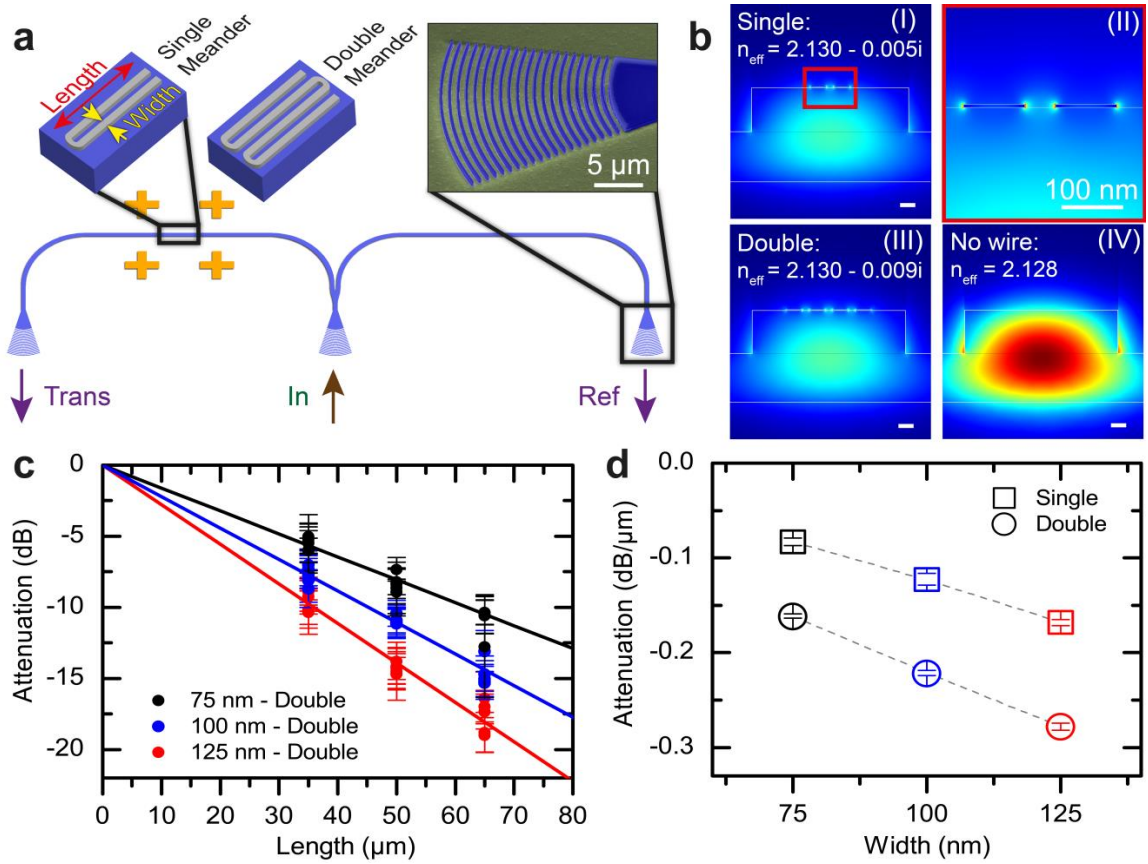

**Figure 2. Absorption characteristics of diamond SNSPDs.** a) Schematic of a photonic circuit for balanced detection of the absorption of a NbN nanowire which is placed on top of the diamond rib waveguide. Inset: False color SEM image of a grating coupler. b) FEM simulation of the fundamental TE-like optical mode showing the electric field norm for a 1 µm wide rib waveguide with a single meander (I and II), with a double meander (III) and without a NbN nanowire (IV). At the NbN nanowire the field is greatly enhanced, as can be seen in the close-up image (II). Scale bars are 100 nm. Subfigure (IV) has a different color scale than the other subfigures in order to make the distribution of the optical mode in the waveguide more visible. c) Measured attenuation of light due to absorption by the NbN nanowires with double meander geometry for different widths (75 nm – 125 nm) and lengths (30 – 70 µm) d) Attenuation

coefficients extracted from linear fits to the measurement data shown in Fig. 2c for single and double meander nanowires of varying wire widths.

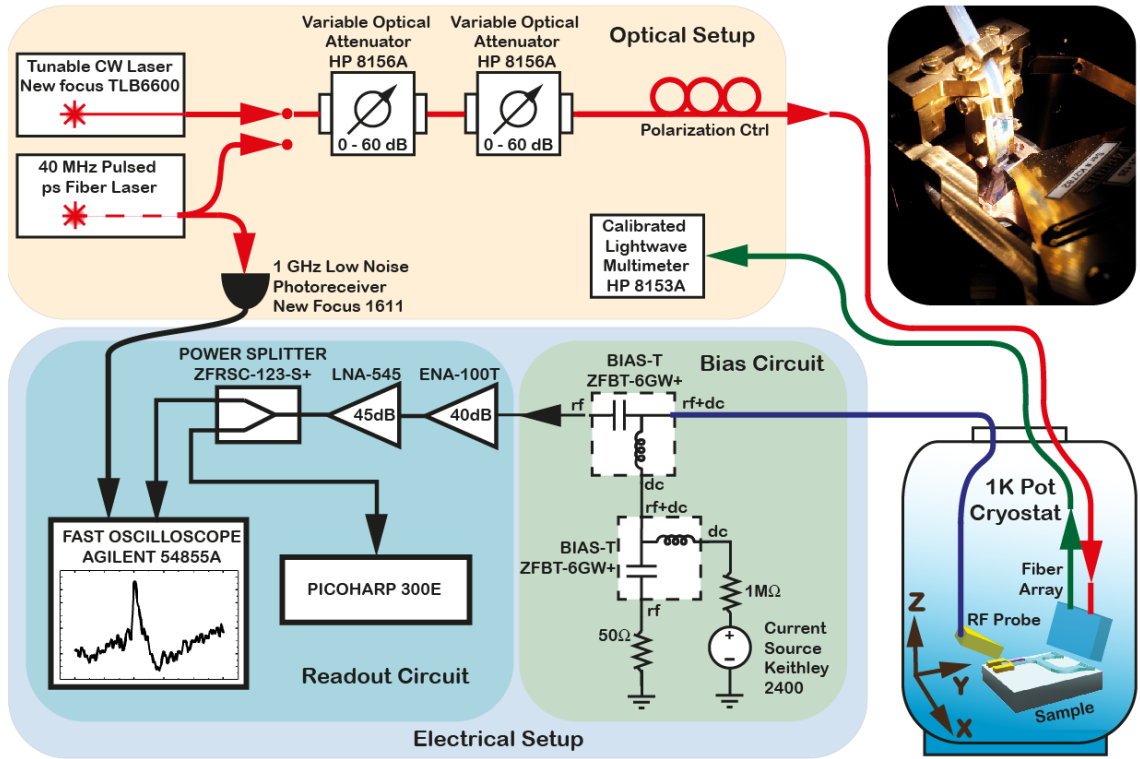

**Figure 3. Cryogenic measurement setup**: The detector sample is mounted inside the liquid helium cryostat at temperatures down to 1K and with optical access (red and green lines) and electrical access (blue line). The optical setup including laser sources, attenuators and reference detectors and the electrical setup including the bias current source for the SNSPDs and the readout electronics are operated at room temperature.

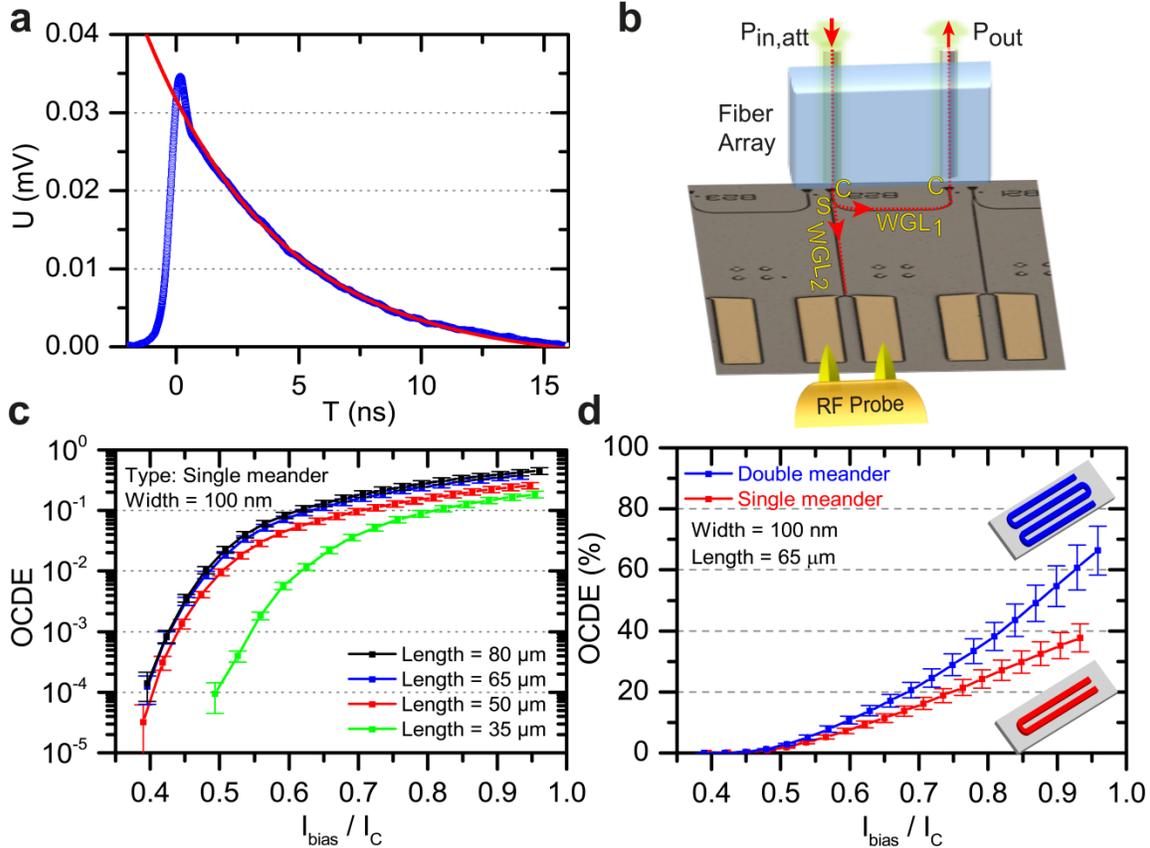

**Figure 4. Characterization of the on-chip detection efficiency.** a) Typical detected pulse shape (blue), resulting from the breakdown and the restoration of superconductivity following the absorption photon in a 35 μm long double meander. The presented data corresponds to the average over 500 detection pulses. An exponential fit to the decay (red) reveals a decay time of 5.1 ns. b) Schematic, showing the fiber array glass tip aligned to the input and reference grating coupler for photon flux monitoring and the electrical RF probe contacted to the metal contact pads of the NbN nanowire detector. c) Dependence of the on-chip detection efficiency on normalized bias current for 100 nm wide single meander wires of varying length, measured at 1.9K. d) Comparison of the on-chip detection efficiency (OCDE) for single and double meander detectors, measured at 1.9K, showing an increased OCDE up to 66%.

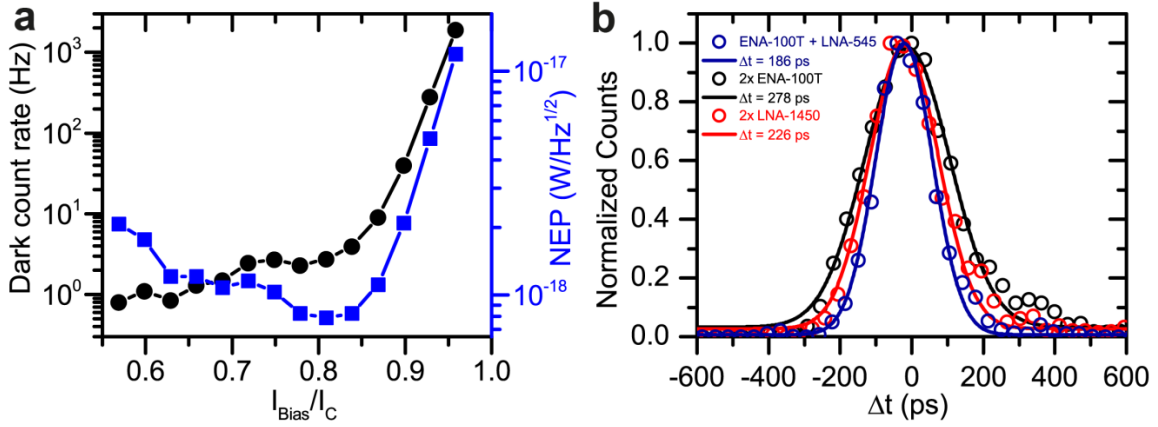

**Figure 5. Dark count performance and timing resolution** a) Dark count rate (black dots) and noise equivalent power (blue squares), measured at 1.9K, for the detector with 66% OCDE (blue data points in Fig. 4d). The maximum dark count rate, when biased at 96% of the critical current is 1.8 kHz. The minimum noise equivalent power, when biased at 81% of the critical current amounts to $7.9 * 10^{-19} \frac{W}{\sqrt{Hz}}$. b) Histograms of the SNSPD timing jitter measured with three different sets of amplifiers. Gaussian fits to the data reveal a FWHM jitter value of 186 ps for the best amplifier combination.